\documentclass[nofootinbib,superscriptaddress,a4paper,twocolumn]{revtex4-1}
\usepackage{geometry}
\usepackage[english]{babel}
\usepackage[latin1]{inputenc}
\usepackage{hyperref}
\usepackage{amsfonts}
\usepackage{amsmath, amssymb, amsfonts, mathrsfs}
\usepackage{graphicx}
\usepackage{xspace}
\usepackage{bbm}
\usepackage{natbib}
\usepackage{multirow}                          
\usepackage[bottom]{footmisc}
\usepackage{array}% http://ctan.org/pkg/array
\usepackage{tabulary}
\usepackage{booktabs}

\AtBeginDocument{
\heavyrulewidth=.08em
\lightrulewidth=.05em
\cmidrulewidth=.03em
\belowrulesep=.65ex
\belowbottomsep=0pt
\aboverulesep=.4ex
\abovetopsep=0pt
\cmidrulesep=\doublerulesep
\cmidrulekern=.5em
\defaultaddspace=.5em
}

\newcommand{\theseus}{{\small T}{\scriptsize HESEUS}\xspace}

\newcommand{\ket}[1]{|#1\rangle}

\newcommand{\cnot}{{\small C}{\scriptsize NOT}\xspace}

\begin{document}

\title{Conceptual understanding through\\ efficient inverse-design of quantum optical experiments}

\author{Mario Krenn}
\email{mario.krenn@utoronto.ca}
\affiliation{Chemical Physics Theory Group, Department of Chemistry, University of Toronto, Canada.}
\affiliation{Department of Computer Science, University of Toronto, Canada.}
\affiliation{Vector Institute for Artificial Intelligence, Toronto, Canada.}
\author{Jakob S. Kottmann}
\affiliation{Chemical Physics Theory Group, Department of Chemistry, University of Toronto, Canada.}
\affiliation{Department of Computer Science, University of Toronto, Canada.}
\author{Nora Tischler}
\affiliation{Centre for Quantum Dynamics, Griffith University, Brisbane, Australia.}
\author{Al\'an Aspuru-Guzik}
\email{alan@aspuru.com}
\affiliation{Chemical Physics Theory Group, Department of Chemistry, University of Toronto, Canada.}
\affiliation{Department of Computer Science, University of Toronto, Canada.}
\affiliation{Vector Institute for Artificial Intelligence, Toronto, Canada.}
\affiliation{Canadian Institute for Advanced Research (CIFAR) Lebovic Fellow, Toronto, Canada}

\begin{abstract}
One crucial question within artificial intelligence research is how this technology can be used to discover \textit{new} scientific concepts and ideas. We present \theseus, an explainable AI algorithm that can contribute to science at a conceptual level. This work entails four significant contributions. (i) We introduce an interpretable representation of quantum optical experiments amenable to algorithmic use. (ii) We develop an inverse-design approach for new quantum experiments, which is orders of magnitudes faster than the best previous methods. (iii) We solve several crucial open questions in quantum optics, which is expected to advance photonic technology. Finally, and most importantly, (iv) the interpretable representation and drastic speedup produce solutions that a human scientist can interpret outright to discover new scientific concepts. We anticipate that \theseus will become an essential tool in quantum optics and photonic hardware, with potential applicability to other quantum physical disciplines. 
\end{abstract}
\date{\today}
\maketitle

\subsection*{Introduction}
Photons are at the core of many quantum technologies that promise advances for imaging applications \cite{moreau2019imaging}, efficient metrological schemes \cite{polino2020photonic}, fundamentally secure communication protocols \cite{flamini2018photonic} as well as simulation \cite{aspuru2012photonic} and computation techniques \cite{aaronson2011computational,peruzzo2014variational,gimeno2015three} that are beyond the capabilities of their classical counterparts. Besides, photons are also among the core players in the experimental investigation of fundamental questions about the local and realistic nature of our universe.

Motivated by these opportunities, recent years have seen dramatic advances in quantum optical technology, which include highly complex operations in integrated photonic chips \cite{carolan2015universal,wang2019integrated,feng2019progress,llewellyn2020chip,lu2020three}, generation of complex multiphotonic entanglement and its application \cite{rubinsztein2016roadmap,wang201818,luo2019quantum}, and the development and application of high-quality deterministic single-photon emitters \cite{wang2017high} and highly efficient photon-number resolving detectors \cite{flamini2018photonic,slussarenko2019photonic}.

To advance technological and fundamental progress further and to enable the exploration of numerous proposed ideas in the laboratories, new experimental concepts and ideas are instrumental. Frequently, however, the design of experimental setups even for well-defined targets is challenging for the intuitions of human experts, and existing systematic schemes (e.g. \cite{vanmeter2007general}) to date only provide solutions for specific experimental scenarios. For that reason, computational design methods for quantum optical experiments have been introduced \cite{krenn2020computer}, in the form of topological search augmented with machine learning \cite{krenn2016automated,zhan2020experimental}, genetic algorithms \cite{knott2016search,nichols2019designing}, active learning approaches \cite{melnikov2018active}, and optimization of parametrized setups \cite{arrazola2019machine}. Unfortunately, due to the complexity and size of the Hilbert space as well as the breadth of quantum optical applications, those algorithms may have severe drawbacks, such as inefficient discovery rates, requirements of a huge amount of training data or specialization on narrow sets of problems. Most importantly, no method so far has shown how to systematically extract scientific ideas, concepts and understanding from the solutions of the computer algorithm. 

Here we demonstrate \theseus, an inverse-design algorithm for quantum optics with highly interpretable representation that allows to scientists to rationalize the solutions quickly. \theseus is generally applicable to discrete-variable quantum optics problems (including post-selected and heralded states, probabilistic and deterministic photon sources), does not need training data, and is orders of magnitude faster than previous comparable approaches. The speed-up allows for the application of topological optimization, which uncovers the conceptual cores underlying the solution. Physicists can then interpret, understand and generalize the underlying ideas and concepts. These advances allows us to apply \theseus to solve several previously open questions about quantum experiments. Concretely, we investigate complex multiphotonic entanglement, the generation of heralded entanglement and complex photonic quantum transformations. In all of these cases, we uncover previously unknown generalizable patterns and new experimental ideas and interpretations. 

Our approach differs significantly from others that try to employ machine learning to extract scientific concepts. The main difference is that these applications so far have been applied to \textit{re}discover previously known concepts \cite{roscher2020explainable}. Examples involve the identification of astronomical concepts such as the heliocentric worldview which has already been considered by Copernicus \cite{iten2020discovering}, the arrow of time and related thermodynamical concepts that were discovered in the 20th century \cite{seif2020machine} or the identification of certain interferometric devices that are used by optical physicists for many years \cite{melnikov2018active}. Those are significant works that indicate great future possibility. However, they come with a grain of salt: It is not clear how much prior knowledge the scientists implicitly use to identify those concepts from the computer algorithms. Therefore, it is unclear how to extend these methods to actual open questions.

In quantum optics, in two works new concepts have been identified \cite{krenn2017entanglement,gao2019computer} using a brute-force computational search algorithm \cite{krenn2016automated}. There, 10.000s of CPU-hours were necessary to arrive at a useful design. Those solutions were represented directed as a sequence of optical elements, which are very unintuitive to interpret conceptually. Moreover, the sequences were highly non-optimized because they emerged through random processes. As a consequence, it required scientists weeks or even months to rationalize the underlying principles (see \cite{krenn2020computer} for more details).

In contrast to those previous approaches, we introduce for the first time an algorithm that produces highly interpretable solutions, which we apply to unsolved problems in science. The discovered solutions allow human scientists to rationalize the new, underlying concepts in quasi-real-time. We explicitly demonstrate this by solving several previously unresolved questions. In all of those cases, we can interpret and understand the underlying design concepts outright. To the best of our knowledge, \theseus is the first algorithm that can provide targeted and systematically new conceptual understanding in a scientific domain. We believe, therefore, that \theseus is an important step towards the goal of interpretable and explainable AI (XAI) in science that will assist human researchers at a conceptual level.

\begin{figure}[t]
\centering
\includegraphics[width=0.42\textwidth]{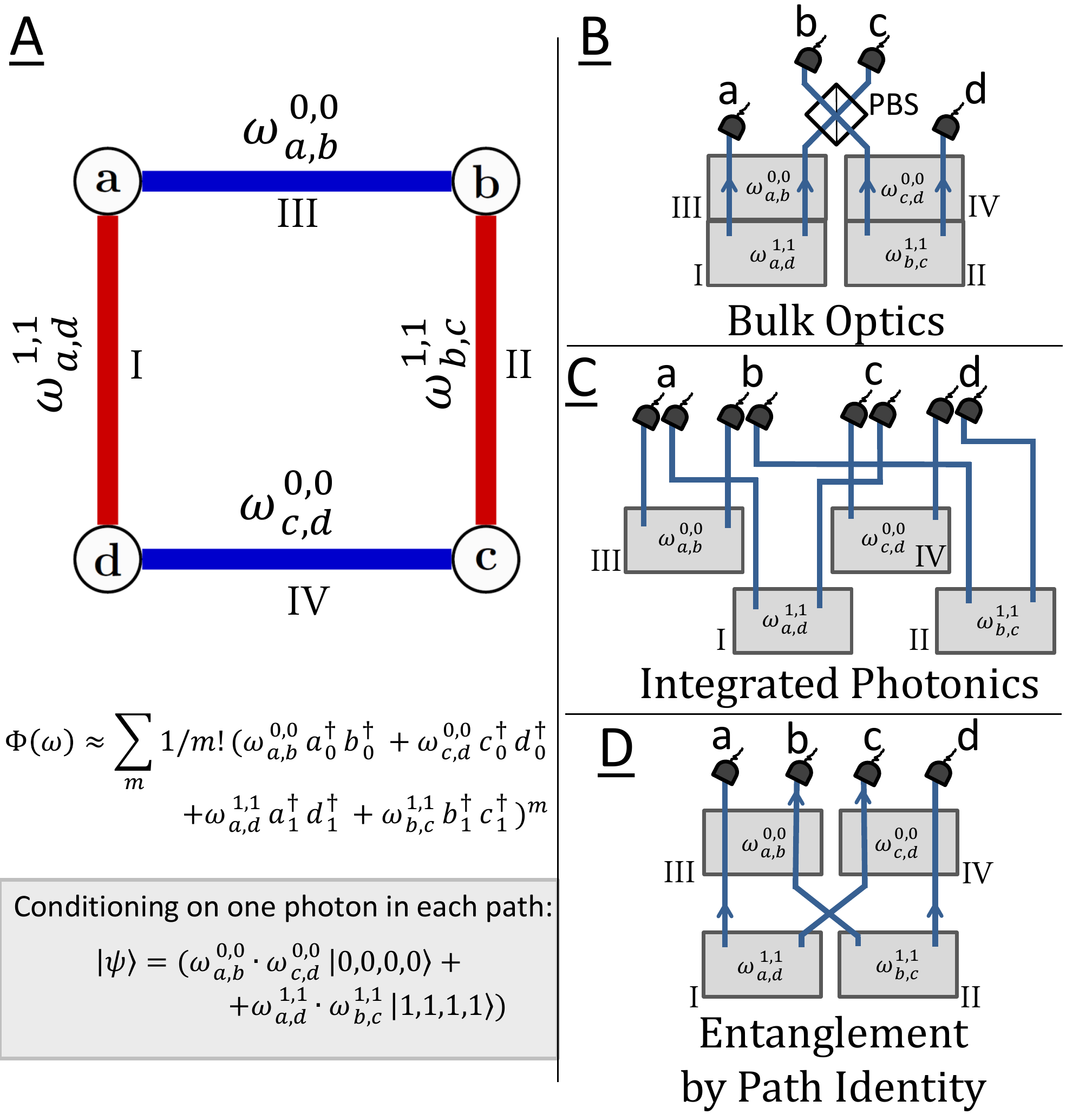}
\caption{A weighted edge-coloured graph as an abstract and efficient representation of the quantum information of a large variety of quantum optics experiments. \textbf{A:} As a specific example, we show a graph with four vertices and four coloured and weighted edges. The vertices $a-d$ correspond to photonic paths, the edges correspond to correlated photon pairs, the edge colours stand for mode numbers, and weights $\omega \in \mathbb{C}$ stand for complex coefficients. Probabilistic sources create the photon pairs (edges). Thus the entire information about the quantum state is represented by $\Phi(\mathbf{\omega})$, with $x^{\dagger}_k$ being a creation operator of a photon in path $x$ with mode number $k$. The information carried in the graph can be translated to different schemes of quantum optical experiments, such as (\textbf{B}) standard bulk optics, for example with path encoding, or (\textbf{C}) polarisation encodings as the carrier of quantum information or (\textbf{D}) Entanglement by Path Identity. The results of the quantum experiments can directly be calculated from the information of the graph. For example, a prominent technique is to condition the state on detecting a photon in each of the four detectors (post-selection). The equivalent formulation in terms of graphs is the sum of all subsets of edges that contain every vertex exactly once. It reduces the example quantum state to two terms. If all weights are equal, the resulting quantum state is a four-qubit GHZ state. Access to $\Phi(\mathbf{\omega})$ allows for the optimisation of non-postselected, heralded and triggered quantum states too, as we show in examples within the manuscript.}
\label{fig:explain}
\end{figure}

\subsection*{Results}

\textbf{Graph Theory--Quantum Experiment Mapping} -- Weighted coloured graphs (explained in Fig.\ref{fig:explain}) can encode the information produced by a photonic quantum experiment involving probabilistic photon-pair sources \cite{krenn2017quantum} and linear optical components \cite{gu2019quantum2}. The vertices correspond to spatial photon paths and edges between vertex $v_1$ and $v_2$ stand for probabilistic photon pairs in path $v_1$ and $v_2$. The edge colour represents the internal mode number of the photons and edge weights $\omega$ stand for amplitudes. Previously, the description was only applicable to post-selected states.

We significantly extend the abstract description of quantum optics experiments as coloured weighted graphs, demonstrating how general quantum optics technology and questions can be raised and solved using the new framework. The extensions allow us for the first time to use the framework of weighted colored graphs for computational design of quantum optical experiments and hardware. 

Specifically, here we introduce a weight function $\Phi(\mathbf{\omega})$ that gives access to the complete information of quantum optical experiments through a  (rather than only post-selected states, as in \cite{krenn2017quantum,gu2019quantum2,gu2019quantum3}), and allows us to generalize the scope of the method significantly. It allows for the description of non-postselected states, triggered and heralded photonic states, states with multiple excitations per mode (such as NOON states) and general quantum transformations. Furthermore, it enables the description of photon-number sensitive and insensitive detectors (which correspond to different projections of $\Phi(\mathbf{\omega})$) and deterministic photon sources such as quantum dots (see SI for details).

Furthermore, we introduce here how graphs can be directly translated to several different schemes of photonic quantum optics, such as standard bulk optics, integrated photonics or entanglement by path identity \cite{krenn2017entanglement}. A given graph can be translated in multiple ways to quantum experimental setups, while each setup corresponds to precisely one graph (more details in SI). These extensions were necessary to use the graph-theoretical description as a tool for the inverse-design of quantum experiments that are feasible in state-of-the-art quantum photonics laboratories.

\begin{figure*}[!t]
\centering
\includegraphics[width=0.85\textwidth]{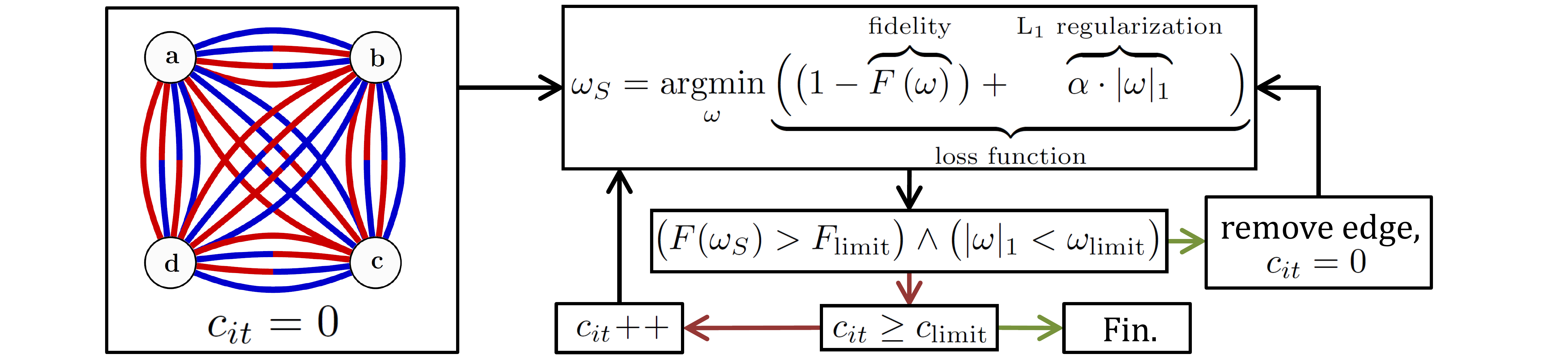}
\caption{\textbf{Algorithm -- \theseus}: The initial graph contains all possible edges between each vertex, leading to $|G|=d^2\frac{n(n-1)}{2}$ edges (with $n$ vertices and $d$ different edge colors), each of them having an independent complex weight $\omega_{v_1,v_2}^{m_1,m_2}$. The main step is a minimization of the loss function, which contains the quantum fidelity in terms of weights of the graph. Additionally, an $L_1$ regularization term controls the magnitude of the weights. If the weights identified by the optimisation, $\omega_S$, lead to fidelities larger than a $F_{\text{limit}}$, and the magnitude of the weights is smaller than $\omega_{\text{limit}}$, then one edge of the graph is removed, and the optimisation continues with the smaller graph. On the other hand, if the criteria are not fulfilled, the same graph is optimised (with different starting conditions) until the discovery of a suitable solution, or the number of iterations exceeds $c_{\text{limit}}$. The result of \theseus is a weighted graph that leads to sufficiently large fidelities, with a small number of edges. This topological optimisation enables the scientific interpretation and understanding of results.} 
\label{fig:algo}
\end{figure*}

\textbf{Graph-based inverse-design of Quantum Experiments} --  
The abstract and general representation of quantum experiments as graphs allows us to find a new method for inverse-designing quantum experiments. The idea is to write an optimisation objective function in terms of weights $\omega$ of the graph. For example, if we aim to find an experimental setups that produces a specific quantum state, the objective function is the fidelity of the state encoded as graph. If we aim to find transformation, then the objective function is the gate fidelity. Importantly, one can use the same technique for more general optimization targets, where neither the quantum state nor the quantum experiments are known beforehand. Examples are quantum metrology, where the objective function would be the Fisher-Information (written in terms of weights $\omega$), or quantum-enhanced imaging technologies, where the objective function could be a signal-to-noise ratio (again, in terms of weights $\omega$).  

The most general quantum state corresponds to a complete graph with all possible multi-coloured weighted edges between each vertex (see Fig.\ref{fig:algo}). As an essential step, we need to construct the objective (e.g. state fidelity) in terms of weights, $F(\omega)$. While the entire quantum state $\Phi(\mathbf{\omega})$ is directly defined by the edge weights, conditioning measurements are commonly used to obtain more intricate states and to overcome the lack of single-photon nonlinearities. Prominent examples for such measurements are conditioning on the simultaneous detection of one photon in each path (I) or conditioning on the detection of ancilla photons (II).

As an example in Fig.\ref{fig:explain}, we show the construction of the fidelity for a 4-photon GHZ state $\ket{GHZ}=1/\sqrt{2}(\ket{0,0,0,0}+\ket{1,1,1,1})_{a-d}$, where $\ket{0}$ and $\ket{1}$ stand for one photon in the internal mode $0$ and $1$ (such as horizontal or vertical polarisation), respectively. The subscript $a$-$d$ means one photon is in each of the four paths $a$,$b$,$c$ and $d$. Under the condition of simultaneous detection (I), the term $\ket{0,0,0,0}$ can be generated by three different subgraphs: two blue horizontal edges, vertical edges or crossed edges. The weight of a subgraph is the product of all its edge weights. The weight of the whole term is the sum of all weights of the subgraphs. Therefore, the weight of $\ket{0,0,0,0}$ is 

\begin{align}
\omega_{\ket{0,0,0,0}}=\omega_{a,b}^{0,0}\omega_{c,d}^{0,0}+\omega_{a,c}^{0,0}\omega_{b,d}^{0,0}+\omega_{a,d}^{0,0}\omega_{b,c}^{0,0}
\end{align} 

In an equivalent way, the amplitude of $\ket{1,1,1,1}$ can be written in terms of $\omega$. As a result, we have

\begin{align}
F(\omega)=\frac{|\omega_{\ket{0,0,0,0}}+\omega_{\ket{1,1,1,1}}|^2}{2\cdot N(\omega)^2}
\label{eq:fidelity}
\end{align}  
where $N(\omega)$ is a normalisation constant of the state emerging from the graph (more details in SI). 

The weights of the graph are optimised by minimising a loss function constructed from the fidelity and an additional L$_1$ regularisation term

\begin{align}
L(\omega)=\left(1-F\left(\omega\right)\right) + \alpha\cdot |\omega|_1
\label{eq:loss}
\end{align} 
with positive real coefficient $\alpha<1$. Inclusion of the L$_1$ regularization term can drive the optimisation towards a solution with smaller amplitudes, thereby opening ways to further reduce the edges of the graph by removing edges with small weights. For optimisation, we use the Broyden-Fletcher-Goldfarb-Shanno algorithm, an iterative gradient-descent method that approximates Hessians to solve non-linear optimization problems. If we identify a solution with $F(\omega)$ above a limit (we use $F_{\text{limit}}=0.95$) and small weights $\omega$ (we use $\omega_{\text{limit}}=1$), we find a suitable experimental setup candidate. At this point, as the loss minimization is fast, we can perform a topological optimisation. We reduce the size of the graph by iteratively removing individual edges. We can choose the edge from a distribution that depends on the magnitude of the weights of the previous solution (with two special cases: choosing entirely randomly, and always choosing the edge with the smallest weight magnitude). The new, smaller graph will be used to minimize the loss function in eq.(\ref{eq:loss}). The topological optimisation reduces the size of the graph iteratively. 

The topological optimisation distills small structures such that human scientists can interpret and understand the underlying physical principles, and use the new knowledge to solve other cases. In many instances, we used these insights to find straight-forward generalizations to infinitely large classes of situations. This is in stark contrast to typical artificial intelligence applications in the natural sciences \cite{carleo2019machine}, where the solution of a parameter optimisation is the final product, without the intention of discovering new scientific ideas (with few recent exceptions that re-discover previously known physical ideas \cite{iten2020discovering,seif2020machine}).

\textbf{Benchmarking} -- We compare the speed of \theseus with previous approaches, using classes of high-dimensional multipartite states called Schmidt-Rank Vectors (SRV) as a benchmark \cite{huber2013structure}. In particular, we aim to discover maximally entangled three-party quantum states of up to ten local dimensions. This task is well understood theoretically, thus it represents a good benchmark. There are 81 unique entangled structures that could be generated using linear optics \cite{gu2019quantum3}. A pure topological circuit search using 150 CPU-hours has discovered 51 out of 81 different states in the set \cite{krenn2016automated}. A reinforcement learning algorithm has identified 17 out of 81 different states, with speed comparable to the topological search algorithm \cite{melnikov2018active}. \theseus discovers 76 different states within 2 hours, where the first 17 are identified within two seconds, and the first 51 states in less than 15 minutes. This results in a speed-up of a factor 600.

We turn to a second benchmarking task; the identification of high-dimensional \cnot gates. A recent study has shown that the identification of the first photonic high-dimensional controlled operation took 150.000 CPU-hours \cite{gao2019computer}. Our algorithm finds a solution that is experimentally quantitatively simpler, within 1 CPU-second. This results in a speedup of a factor $10^8$. We come back to this example in Fig.\ref{fig:2dCNOT}.

\begin{figure}[b]
\centering
\includegraphics[width=0.49\textwidth]{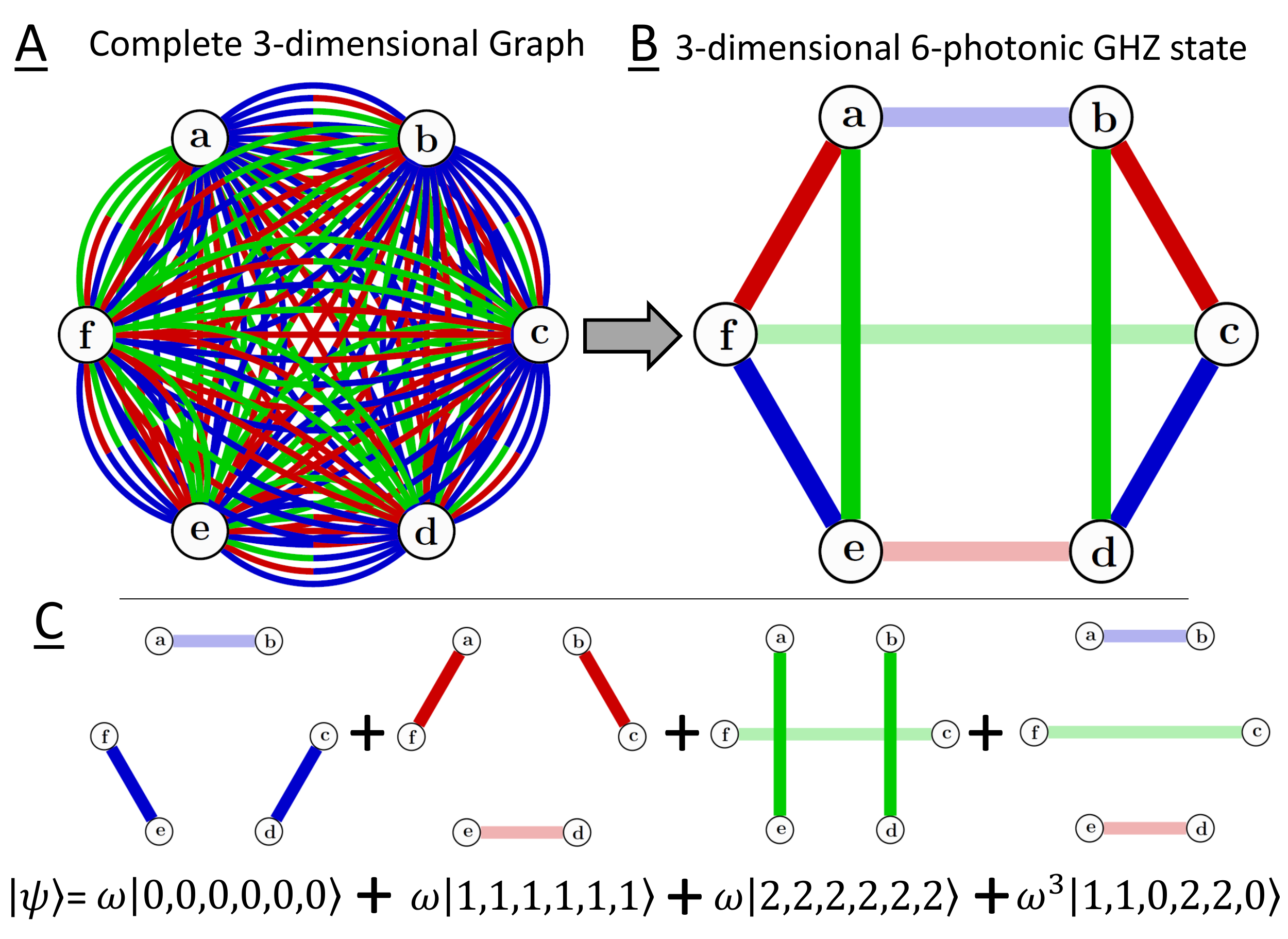}
\caption{\textbf{Finding a 3-dimensional 6-photonic GHZ state.} \textbf{A:} The initial state is a complete graph of six vertices and three colours. Each pair of vertices is connected by nine edges, which stand for all nine possibilities (blue, red, green stands for modes 0,1,2, respectively). A bi-coloured edge stands for a photon pair with different mode numbers. For example, a blue-red edge between vertex $a$ and $b$ stands for a photon pair with one photon in path $a$ with mode number $0$, and one photon in $b$ with mode number $1$, i.e. $a^{\dagger}_0 b^{\dagger}_1$. In total, this corresponds to 135 edges. \textbf{B:} The optimized graph for a 6-photonic 3-dimensional GHZ state. While it has been shown that such a state cannot be created with perfect fidelity \cite{krenn2017quantum} with linear optics and probabilistic photon-pair sources (without additional photons), \theseus found a solution where the fidelity scales with $F\approx 1 - O(\omega^4)$ with the overall counts $C$ scaling as $C \approx O(\omega^2)$, which is experimentally feasible. \textbf{A:} The concept of the solution can be interpreted in the context of graph-theoretical results and can be immediately generalized by human scientists.} 
\label{fig:6n3dGHZ}
\end{figure}

\begin{figure*}[t]
\centering
\includegraphics[width=1\textwidth]{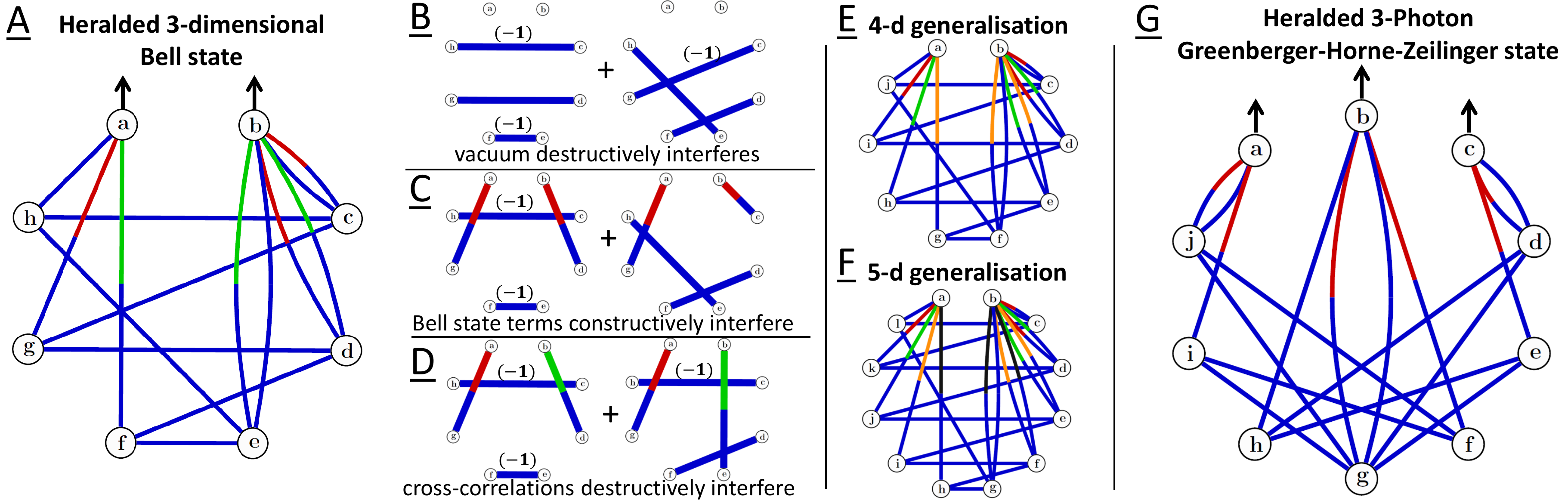}
\caption{\textbf{Heralded entangled states.} \textbf{A:} Optimized graph for a heralded 3-dimensionally entangled Bell state with photon in $a$ and $b$ containing a Bell state if detectors in $c$-$h$ click, which requires eight photons in total. The crucial insight is the destructive interference of the undesired heralded vacuum term. \textbf{B:} Each of the two subgraphs herald the vacuum. The overall weight of the two subgraphs differs only in sign leading to destructive interference. \textbf{C:} With the same phase setting, all terms of the Bell state constructively interfere, such as $\ket{1,1}$ , and all cross-correlation terms cancel, such as $\ket{1,2}$ in \textbf{D}. This solution can immediately be generalised to arbitrary high-dimensional Bell states, see \textbf{E} and \textbf{F}. \textbf{G:} Solution for a heralded 2-dimensional GHZ state in $a$,$b$,$c$, with triggers in $d-j$, requiring 10 photons. The algorithm identifies a highly symmetric solution that destructively interferes 8-photon contributions which heralds the trigger} 
\label{fig:Heralded}
\end{figure*}

\textbf{Scientific Discovery and Understanding} -- 
The improvement in speed shows that \theseus is ready to go beyond benchmarks, and be applied to the discovery of new scientific targets and to the development of new scientific insights and understanding. Scientific understanding is essential to the epistemic aims of science \cite{de2005contextual}, but rarely addressed in applications of artificial intelligence to the natural sciences. In the philosophy science, pragmatic criteria have been found for \textit{scientific understanding}, in particular by de Regt's award-winning work \cite{de2005contextual, de2017understanding}. He describes that scientists can understand a phenomenon \textit{if they can recognise qualitatively characteristic consequences without performing exact calculations}. We connect this criterion to our discoveries: We discover the first high-dimensional six-photonic GHZ states, which have been conjectured to be not constructible with linear optics. We can understand the underlying concept and use it to construct a simple method to generate high-dimensional GHZ states with an arbitrary number of photons. Furthermore, we discover the first solutions of heralded three-dimensional Bell states. We also understand the underlying concept, which, among others, contains an idea to destructively interfere vacuum terms. We generalise the concept to arbitrary-dimensional Bell states -- without additional calculations. Similarly, we discover setups for heralded GHZ states that need fewer resources than methods proposed in the literature, which could form the building blocks for photonic quantum computation \cite{gimeno2015three}. We furthermore apply \theseus on multiphotonic transformations. We find a new way to interpret and construct photonic qubit operations such as \cnot gates. Similarly, we discover high-dimensional \cnot operations that need quantitatively less resources than methods proposed in the literature. Connecting to de Regt's criterion, our algorithm has been the source of scientific understanding for multiple instances.
 
\textit{High-dimensional GHZ states} -- 
A d-dimensional n-partite GHZ quantum state is written as

\begin{align}
\ket{\psi}=\frac{1}{\sqrt{d}}\sum_{i=0}^{d-1}\ket{\underbrace{i,i,i,\dots}_{\text{n times}}}.
\label{eq:GHZdn}
\end{align} 
These states are studied in the interplay between quantum and local-realistic theories \cite{lawrence2014rotational}, and have recently found potential applications in quantum communication tasks \cite{hu2020experimental}.

Graph-theoretical arguments show that perfect high-dimensional GHZ states can be generated only for 4-photon states \cite{krenn2017quantum}, because terms in addition to those in eq.(\ref{eq:GHZdn}) necessarily emerge. Using \theseus, we discover the first example that circumvents the no-go theorem, see Fig.\ref{fig:6n3dGHZ}. The algorithm identifies solutions with fidelities arbitrarily close to one, by adjusting the edge weights such that unwanted terms have arbitrarily small weights (albeit at the expense of lower count rates). The solution can  immediately be generalized to GHZ states (and other states) with higher dimensions and a larger number of particles, by identifying subgraphs of additional terms who's edges are multiplied with $\omega<1$.

No further computations or optimisations are necessary, demonstrating that we achieved \textit{scientific understanding} based on a computational optimisation in the appropriate representation of the problem at hand.

\begin{figure*}[t]
\centering
\includegraphics[width=0.87\textwidth]{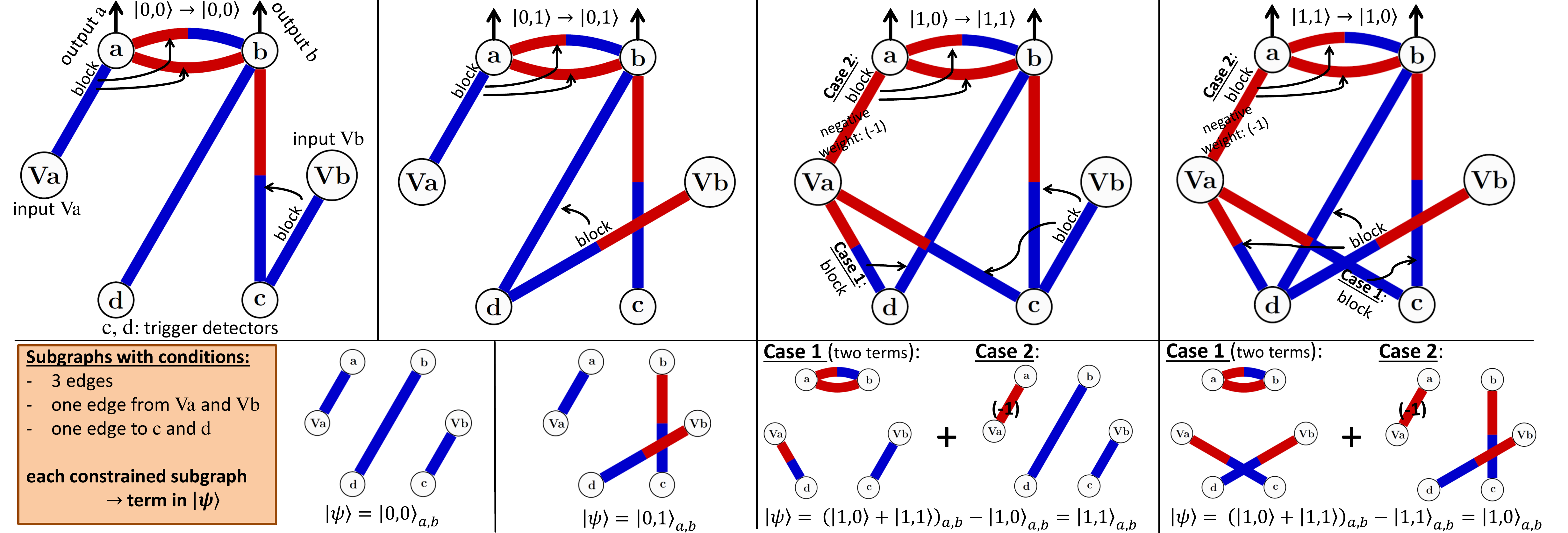}
\caption{\textbf{Qubit \cnot Transformation}: Two input photons (denoted as Va and Vb) undergo a \cnot operation, and output in path $a$ and $b$, conditioned on the simultaneous detection of one photon in each of the trigger paths $c$ and $d$. This example goes beyond state generation and shows how the same framework allows identification of quantum transformations. We introduce \textit{virtual} vertices Va and Vb, which are interpreted as incoming photons. Edges between a virtual vertex and vertices $a$-$d$ (in all graphs together) represent unitary transformations of the incoming photon. For example, if Vb=$\ket{0}$, the photon goes to path $c$, while for Vb=$\ket{1}$ it goes to path $d$. The \cnot consists of four individual transformations (one for each of the inputs $\ket{0,0}$,$\ket{0,1}$,$\ket{1,0}$,$\ket{1,1}$). Each transformation stands for one graph and the subgraph of vertices $a$-$d$ has to stay constant for each transformation. The four graphs in the upper row are the solution of an inverse-design for a two-qubit \cnot. The quantum state in the output of $a$,$b$ (after conditioning on the trigger detectors $c$,$d$) is governed by all subgraphs that fulfil the following conditions (see orange inset): Contains three edges (two edges from incoming photons, and one ancillary photon pair);each of the virtual vertices Va and Vb is contained in one edge (that symbolizes that one photon is entering the setup), and both $c$ and $d$ are contained in an edge (such that the two triggers detect a photon in path $c$ and $d$). The solution can be conveniently interpreted: No vertex can have two incoming edges (as follows from the three conditions). Therefore, an edge involving Va or Vb block all edges at the other vertex of the edge, which significantly simplifies the interpretation of the graphs. The resulting terms are written in the lower row. From the solution, we discover an interesting concept: If Va=$\ket{0}$, the edge involving Vb \textit{chooses} the outgoing term by blocking the appropriate edge. However, if Va=$\ket{1}$, the double edge between $a$ and $b$ is active -- as the weight $\omega_{Va,a}^{1,1}=(-1)$, Vb chooses the term that will destructively interfere. The idea of having one virtual vertex choosing the terms can be generalized to more complex multi-qubit transformations.}      
\label{fig:2dCNOT}
\end{figure*}

\textit{Heralded photonic entangled states} -- 
The next targets we address are heralded entangled photonic states. Standard sources of photonic entanglement such as spontaneous parametric down-conversion or spontaneous four-wave mixing, are entirely probabilistic \cite{wang2019integrated}. That means photons are produced at random times, and only after the detection of the photon state, one knows that they have been created. The generation of heralded states would allow for event-ready schemes, which are essential in photonic quantum computation \cite{gimeno2015three,rudolph2017optimistic}. Experimentally, two-dimensional Bell states have been generated conditioned on the detection of photons in four trigger detectors \cite{barz2010heralded, wagenknecht2010experimental}. However, higher-dimensional generalizations implementations are missing. A major challenge in creating heralded states are cases where all trigger detectors see a photon, but no photons emerge from the setup. Those cases, where the triggers herald a vacuum term, usually have significantly higher probability of happening than the correct heralded Bell state because fewer pair creation events need to occure simultaneously. 

\theseus identifies experiments for heralded 3-dimensional Bell state, see Fig.\ref{fig:Heralded}A. The setup requires four photon-pair events simultaneously, which is well within today's experimental capabilities \cite{zhong201812}. The solution contains a remarkable \textit{idea} that not been explored before: The destructive interference of the triggered vacuum term, see Fig.\ref{fig:Heralded}B. Creating the possibility of two heralded vacuum outputs and assigning their amplitudes opposite signs leads to their cancellation.

Furthermore, each of the two subgraphs that lead to a vacuum term in Fig.\ref{fig:Heralded}B forms the basis of a 3-dimensional Bell-state which constructively interfere while all cross-correlations destructively interfere. More information is provided in the SI. Higher-order events and cases where multiple photons are detected in the trigger detectors can be reduced to arbitrarily low probabilities by adjusting the weights of the edges. Assuming a standard pump laser with 80 MHz repetition rate, the expected count rate to reach a fidelity that guarantees genuine 3-dimensional entanglement, i.e. $F>2/3$, is on the order of ten per second (for details see SI). The concepts used by \theseus, in particular the cancellation of vacuum, can be immediatly generalized to other cases, for example, to arbitrary high-dimensional Bell states, see Fig.\ref{fig:Heralded}E-F. 

Next, we use \theseus to find heralded multi-photonic states which have been proposed a decade ago, but never experimentally implemented due to their experimental requirements \cite{walther2007heralded,niu2009heralded}. Heralded GHZ states provide the resources for definite demonstration of deterministic violations of local-realistic worldviews \cite{erven2014experimental} and are among the most promising building blocks for photonic quantum computation \cite{gimeno2015three,rudolph2017optimistic}. We find an experimental configuration, which requires fewer resources and which is within reach of experimental capabilities, see Fig.\ref{fig:Heralded}G. The solution is highly symmetric, and uses a very similar concept to avoid lower-order contributions as the solution of the Bell state. In this case however, the problematic lower-order event create single-photon outputs. The strategy, again, is to generate two subgraphs for each single-photon output with opposite phase which destructively interfere (see SI).

\textit{Photonic Controlled-Gates} -- 
Finally, we demonstrate the usage of \theseus to photonic quantum transformations, which are essential elements for photonic quantum simulation \cite{aspuru2012photonic} and computation schemes \cite{rudolph2017optimistic,flamini2018photonic,slussarenko2019photonic}. In Fig.\ref{fig:2dCNOT} we introduce virtual vertices that represent input photons, and optimize multiple dependent graphs simultaneously that represent different states of the transformation. Interestingly, exactly this concept has been the core of one of the first photonic \cnot experiments \cite{gasparoni2004realization}, which gives a new interpretation for a 16-year-old experiment (see SI for details). 

We apply \theseus to find high-dimensional quantum transformations, which have been discussed in the context of resource-efficient quantum computation algorithms \cite{bocharov2017factoring, jafarzadeh2019randomized}. The solution follows similar concepts as the two-dimensional case, and requires fewer experimental resources than \cite{gao2019computer}, for details see SI.

\subsection*{Discussion}
We presented the algorithm \theseus for the inverse-design of quantum optical experiments, which is based on an abstract physics-inspired representation. We use it to discover several previously unknown experimental configurations of quantum states and transformations in the challenging high-dimensional and multi-photonic regime, such as generation of high-dimensional GHZ states, heralded entangled quantum states, high-dimensional controlled operations. Those experimental setups are within reach of modern photonic technology and could lead to fascinating experimental investigations of fundamental questions and technological advances. \theseus can immediately be applied to discover a multitude of other targets in experimental quantum optics, such as tools to enable silicon-photonics quantum computation \cite{rudolph2017optimistic} or highly efficient, low-noise quantum entanglement sources \cite{zhong201812}. It can also directly be applied to situations where the target state is not known beforehand, such as for applications in quantum metrology \cite{polino2020photonic} or in quantum-enhanced microscopes and telescopes \cite{parazzoli2016enhanced,rafsanjani2017quantum}. In general, the internal representation is directly connected to creation and annihilation operators, which are universally used in quantum physics, thus \theseus can further be generalized to a much larger scope. 

One of the main features is the possibility to extract scientific understanding from the computer-inspired designs. That was made possible by a topological optimisation that reduces the solutions to conceptual cores. Those minimal topologies allow for the interpretation and generalizations of the discovered solution, without performing additional calculations. This is in accord with criteria from the philosophy of science that argue that scientific understanding is connected with the skill to use concepts fruitfully, \textit{without exact calculations}. Hence, in a broader sense, we argue that the ability of our algorithm goes beyond simple optimisation, and enters the realm of providing scientific insights and allowing for scientific understanding. Thereby, it directly contributes to scientific, explainable AI (XAI), and in general, to the essential aim of science.

\section*{Acknowledgements}
This work was supported by the Google Focused Award on Quantum Computing, the Industrial Research Chair Program of Canada and by the U.S. Department of Energy, Office of Science, Office of Advanced Scientific Computing Research, Quantum Algorithm Teams Program. A. A.-G. acknowledges generous support from the Canada 150 Research Chair Program, Tata Steel, Anders G. Fr{\o}seth, and the Office of Naval Research. M.K. acknowledges support from the Austrian Science Fund (FWF) through the Erwin Schr\"odinger fellowship No. J4309. N.T. acknowledges support by the Griffith University Postdoctoral Fellowship Scheme.
\bibliographystyle{unsrt}
\bibliography{refs}

\renewcommand{\figurename}{FigS.}

\clearpage

\widetext
\begin{center}
\Large{Supplementary Information:}\\ 
\textbf{\Large Conceptual understanding through\\ }
\textbf{\Large efficient inverse-design of quantum optical experiments\\ }
Mario Krenn, Jakob Kottmann, Nora Tischler, Al\'an Aspuru-Guzik
\end{center}

\onecolumngrid

\section{Basic Elements of a Graph as Experimental Building Blocks}
\begin{figure}[h]
\centering
\includegraphics[width=0.8\textwidth]{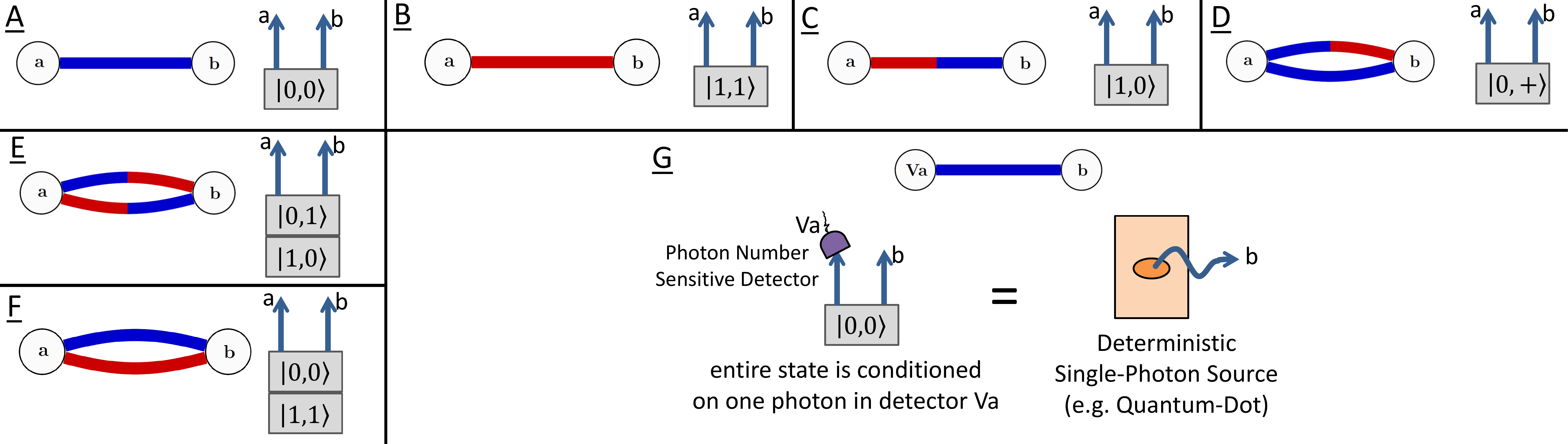}
\caption{Graph to experiment translation for individual edges.}
\label{figSI:GraphExperiment0}
\end{figure}
Designing quantum optical experiments using the abstract notation of graphs is possible because we found translations of graphs into several different experimental schemes. Edges between vertices $a$ and $b$ are translated to probabilistic photon sources, see FigS.\ref{figSI:GraphExperiment0}A-F. Edge colours correspond to mode numbers. Multi-edges correspond to superposition or entanglement, and can be created with standard photonic technologies, for example, cross-crystal sources \cite{hardy1992source, kwiat1995new}. A deterministic single-photon source emitting in path $b$ can be understood as an edge between a vertex $b$ and a virtual vertex Va, FigS.\ref{figSI:GraphExperiment0}G. For each term in the resulting quantum state, every virtual vertices always need to have exactly one incoming edge. This is conceptually equivalent to the situation of a probabilistic photon-pair source, where the whole state is conditioned on the detection of one photon using a photon number sensitive detector in path Va. 
\begin{figure}[b]
\centering
\includegraphics[width=0.8\textwidth]{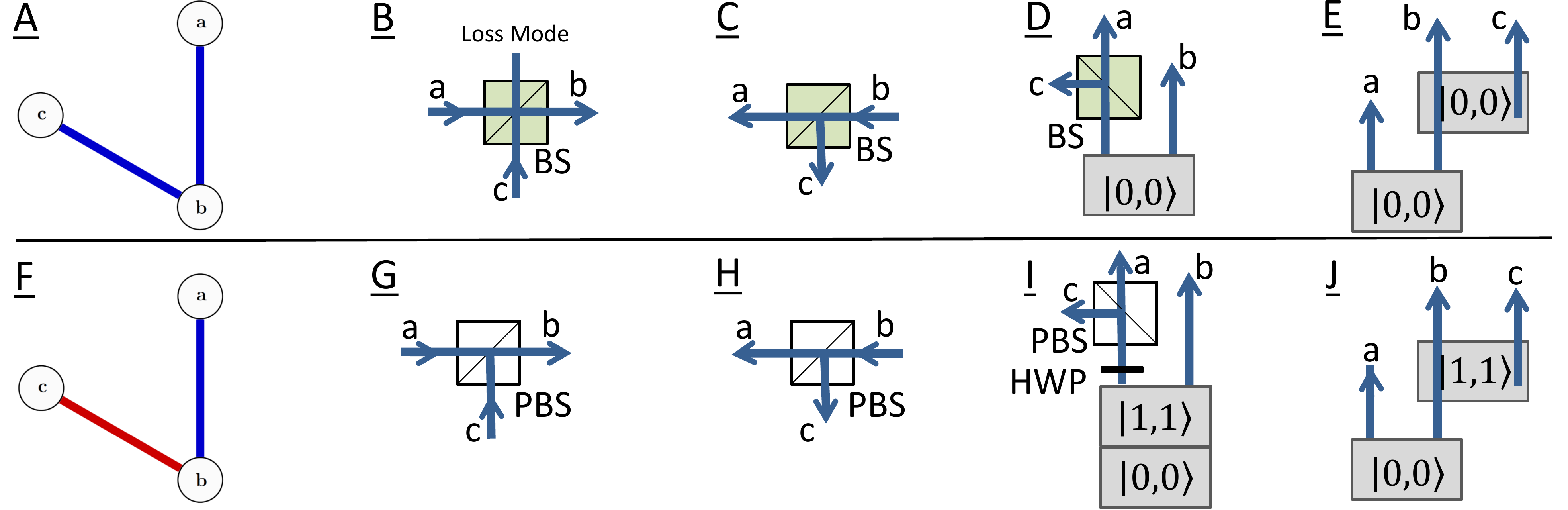}
\caption{Graph to experiment translation. BS: beam splitter, PBS: polarizing beam splitter, HWP: half-wave plate}
\label{figSI:GraphExperiment}
\end{figure}

Edges can be merged at one vertex in several different ways, see FigS.\ref{figSI:GraphExperiment}. If the edges have the same colour (FigS.\ref{figSI:GraphExperiment}A), the corresponding photons have the same mode number. In that case, the edges can be merged with probabilistic beam splitters (green squares, FigS.\ref{figSI:GraphExperiment}B-D) or by creating them directly with path identified photon-pair sources (for instance, SPDC crystals, FigS.\ref{figSI:GraphExperiment}E).

If the edges have different colour (FigS.\ref{figSI:GraphExperiment}F), the corresponding photons have different mode numbers. In that case, the edges can be merged losslessly with mode-dependent beam splitters (so-called multiplexing or de-multiplexing); white squares, for example, polarizing beam splitters if the degree of freedom is photonic polarisation (FigS.\ref{figSI:GraphExperiment}G-I). The edges could also be created by path identified photon-pair sources (for instance, SPDC crystals, FigS.\ref{figSI:GraphExperiment}J. Other probabilistic photon sources, such as lasers as probabilistic single-photon sources, can be added by exploiting hypergraph structures \cite{gu2020quantum}.

With the ability to create independent edges, and merge edges, all types of graphs can be translated to experimental setups. Appropriate phase shifters can manipulate the phases of edge weights. Additionally, amplitudes can be manipulated by pump power for SPDC crystals, splitting ratios that are set by half-wave plates, or absorptive filters. Collinear photon pair sources, that produce two photons in the same path, can be described with loops (an edge that connects a vertex to itself).

\section{Normalization of Quantum States}

\begin{figure}[h]
\centering
\includegraphics[width=0.75\textwidth]{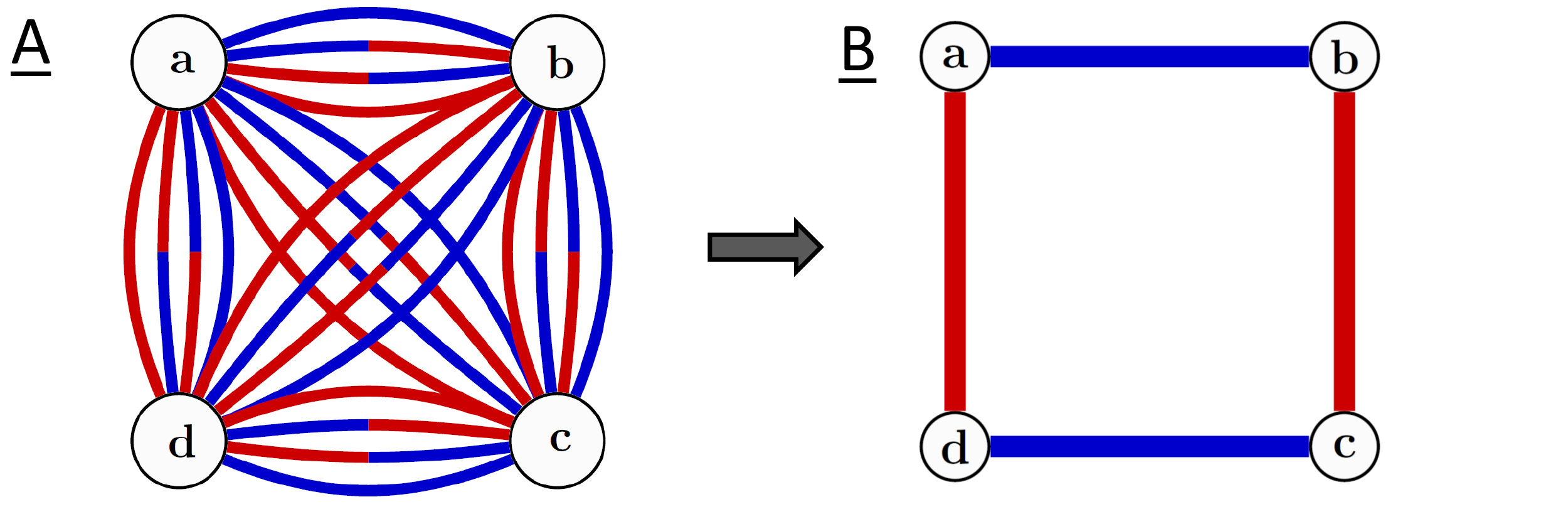}
\caption{\textbf{A:} A complete graph with four edges between each pair of vertices represents all possible correlations in a locally 2-dimensional system. \textbf{B:} In the optimization for a post-selected GHZ, it is reduced to a cycle graph.}
\label{figSI:Complete2d}
\end{figure}

We show how the state of a complete 2-coloured graph with four vertices can be written using the weight function $\Phi(\omega)$ of the graph in FigS.\ref{figSI:Complete2d}A. It can be represented in terms of creation operators as 

\begin{align}
\Phi(\omega) \approx \sum_n \frac{1}{n!}
\left(\sum_{
\substack{
            x,y \in \{a,b,c,d\}\\
            x < y }
}
\sum_{c_1,c_2 \in \{0,1\} } \omega_{x,y}^{c_1,c_2} x_{c_1}^{\dagger} y_{c_2}^{\dagger} + h.c. \right)^n.
\label{SI:phiomega}
\end{align} 
If we are conditioning the state on one photon in each detector, it reduces to

\begin{align}
\ket{\psi}=\frac{1}{N(\omega)} \sum_{i,j,k,l \in \{0,1\}} \omega_{\ket{i,j,k,l}} \ket{i,j,k,l}
\end{align}
with the edge weights
\begin{align}
\omega_{\ket{i,j,k,l}} &= \omega_{a,b}^{i,j}\cdot \omega_{c,d}^{k,l} + \omega_{a,c}^{i,k}\cdot \omega_{b,d}^{j,l} + \omega_{a,d}^{i,l}\cdot \omega_{b,c}^{j,k}
\end{align} 
and the normalization constant
\begin{align}
N(\omega) &= \sqrt{\sum_{i,j,k,l \in \{0,1\}} |\omega_{\ket{i,j,k,l} }|^2}.
\end{align}

The objective of the optimization is to find $\omega_{x,y}^{i,j} \in \mathbb{C}$ that minimize the loss function, and subsequently finding solutions with a large number of edge weights being zero. The information about higher-order contributions to the state, which results in experimentally reduced quantum fidelities, is encoded within the weight function $\Phi(\omega)$. Therefore, higher-order contributions could be directly accounted for within the optimization procedure. More details about the approximations in eq.(\ref{SI:phiomega}) can be found in \cite{zou1991induced,walborn2010spatial}.

\section{Heralded Bell State}
\begin{figure}[h]
\centering
\includegraphics[width=0.75\textwidth]{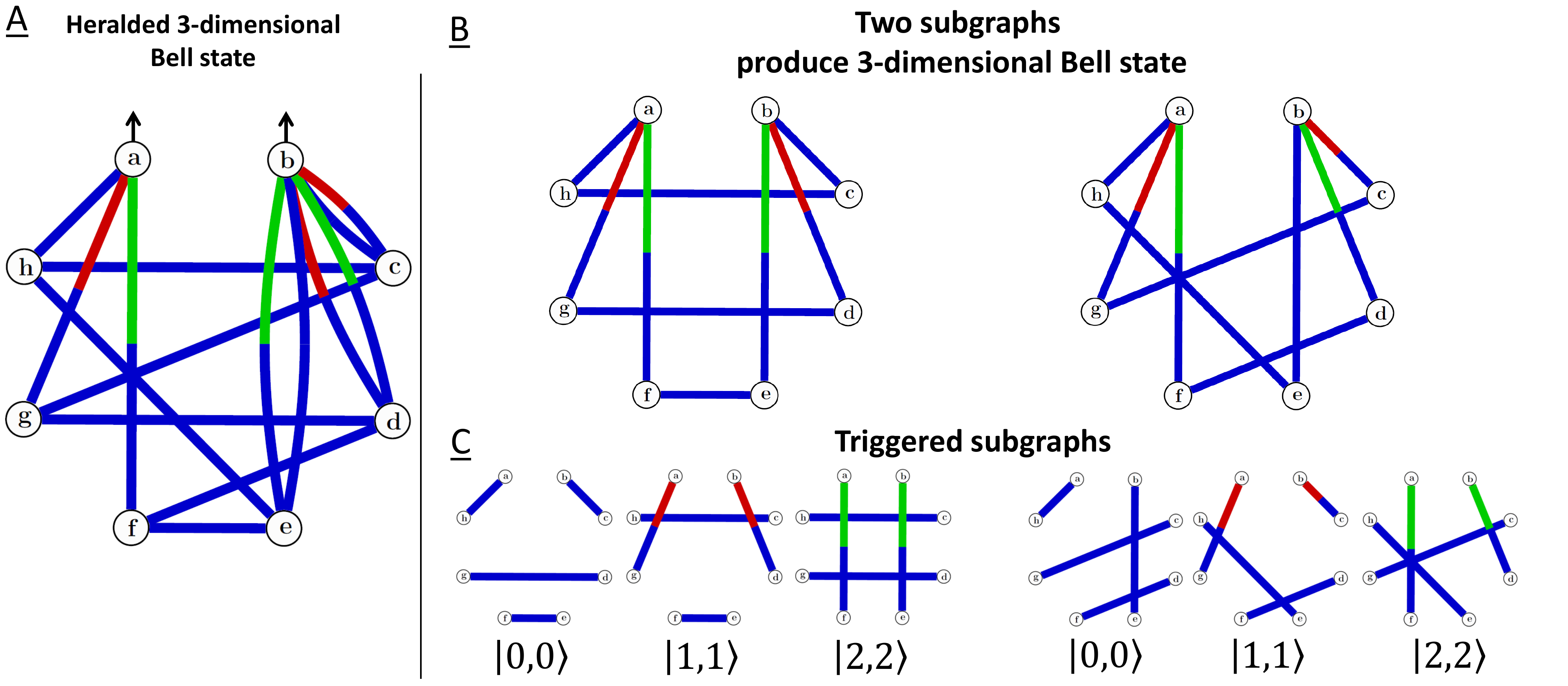}
\caption{The solution for the 3-dimensional Bell state contains two subgraphs, each of them creating individually a 3-dimensional Bell state.}
\label{figSI:Bell1}
\end{figure}

In FigS.\ref{figSI:Bell1}, we show that the solution for the 3-dimensional Bell state contains two subgraphs, each of them creates a 3-dimensional Bell state individually. The vacuum terms of the two subgraphs cancel (as described in the main text). Each of the two subgraphs can be understood individually, and follows the same concept: Every edge from the output modes $a$ and $b$ is connected to one individual ancilla vertex $c-h$. Three edges furthermore connect the ancilla vertices. Each of those edges connects vertices with the same colour of the incoming edge from $a$ and $b$. For example, the left subgraph in FigS.\ref{figSI:Bell1}B has an edge which connects $d$ and $g$ as both of them have an incoming edge with the same colour, red. In that way, if four photon pairs created, only photon pairs with the same edge colour, i.e. mode number, can be created, as seen in FigS.\ref{figSI:Bell1}C.

Cross-correlations, which can occur by combining the two subgraphs in FigS.\ref{figSI:Bell1}B are destructively interfered in the same way as the vacuum with the appropriate setting of the phases of weights, as explained in the main text.

The fidelity can be arbitrary close to one, by adjusting the weights of the edges. In the most straightforward setting, all edges that are connected to $a$ or $b$ have the same weight $v$, while all edges connecting ancilla vertices $c-h$ have weight $w$ (with phases as shown in the main text). In this way, the heralded state can be written as

\begin{align}
\ket{\psi}=2v^2w^2\left(\ket{0,0}-\ket{1,1}-\ket{2,2}\right)_{a,b} + v w^3 \ket{\phi}_{\textnormal{one photon}} + w^4 \ket{\phi}_{\textnormal{zero photons}} + \mathcal{O}\left(\textnormal{higher orders}\right)
\end{align}  

where $\ket{\phi}_{\textnormal{one photon}}$ stands for combinations where three ancilla photon pairs and one pair containing an ancilla photon and an output photon are produced. The state $\ket{\phi}_{\textnormal{zero photon}}$ are cases where four ancillary photon pairs are created. Both of those terms can be reduced by making $w$ smaller than $v$. The term $\mathcal{O}(\textnormal{higher orders})$ correspond to cases with five or more photon pairs produced, which can be reduced by having $v$ and $w$ smaller than one. 

We calculate the fidelity and expected count rates for various settings of weights $v$ and $w$ in Tabel \ref{tab:3dBell}, calculated up to sixth order of SPDC, and not taking into account any losses or detector inefficiency.

\setlength{\tabcolsep}{3pt}
\begin{table*}[t]
\centering
  \begin{tabular}{ c c |  c |  c }
    \toprule
     \textbf{$v$} & \textbf{$w$} & \textbf{fidelity} & \textbf{count rate}\\    
    \midrule
    0.16 & 0.07 & 2/3 & 18.8 Hz \\ 
    0.125 & 0.048 & 0.75 & 1.5 Hz \\     
    0.1 & 0.035 & 0.8 & 0.8 Hz \\  
    0.0820219 & 0.0240018 & 0.85 & 65 per hour\\       
    0.0576405 & 0.0139269 & 0.9 & 1.8 per hour\\          
    \bottomrule
  \end{tabular}
  \caption{Fidelity and count rates for heralded 3-dimensional Bell states.}
\label{tab:3dBell}  
\end{table*}
\begin{figure}[h]
\centering
\includegraphics[width=0.75\textwidth]{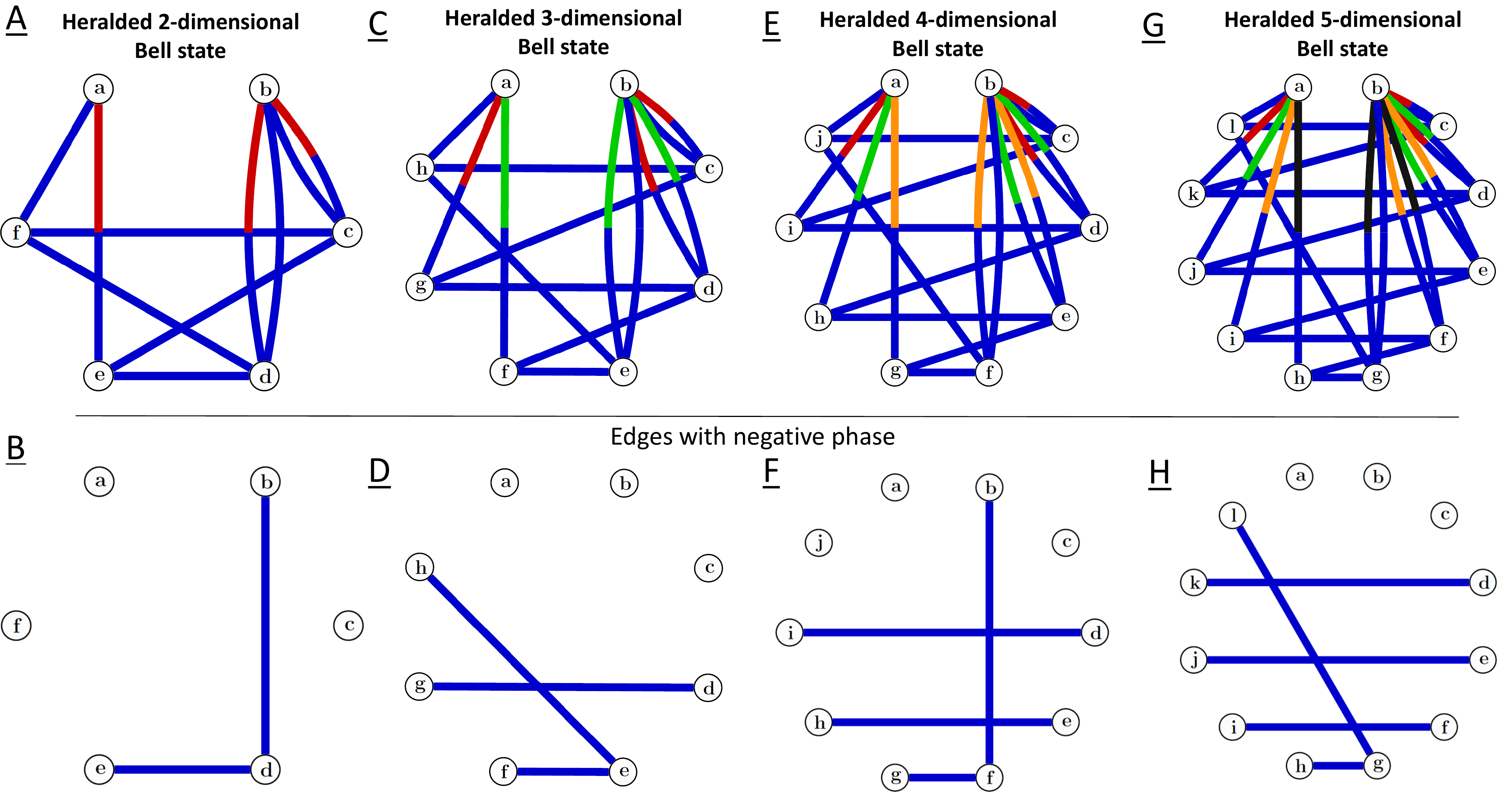}
\caption{Heralded high-dimensional Bell State, and the corresponding phase setting. We identifies the 3-dimensional case with \theseus, and generalized to arbitrary dimensions by understanding the involved concepts and ideas.}
\label{figSI:Bell2}
\end{figure}

The concepts used in the 3-dimensional case can be immediatly generalized to higher-dimensional Bell states. In FigS.\ref{figSI:Bell2}, we show the solutions for 2-dimensional to 5-dimensional Bell states with their corresponding phase settings.

\section{Heralded GHZ State}

More than ten years ago, schemes for heralded GHZ states have been proposed \cite{walther2007heralded,niu2009heralded}, which require experimentally significantly more resources and have therefore not yet became practical. In particular, the 3-photon GHZ proposal by Walther et al. \cite{walther2007heralded,niu2009heralded} requires 12 photons (nine ancillary photons that herald a GHZ state). The proposal by Niu et al., \cite{niu2009heralded} requires ten photons (seven ancillary photons), but further requires close to perfectly efficient, photon-number-sensitive detectors for heralding paths, as they need to distinguish between the arrival of one and two photons. In contrast, our proposal requires only ten photons and non-photon number resolving detectors -- which is feasible in state-of-the-art photonic laboratories.

\section{Experimental 2-qubit CNOT}

\begin{figure}[t]
\centering
\includegraphics[width=0.85\textwidth]{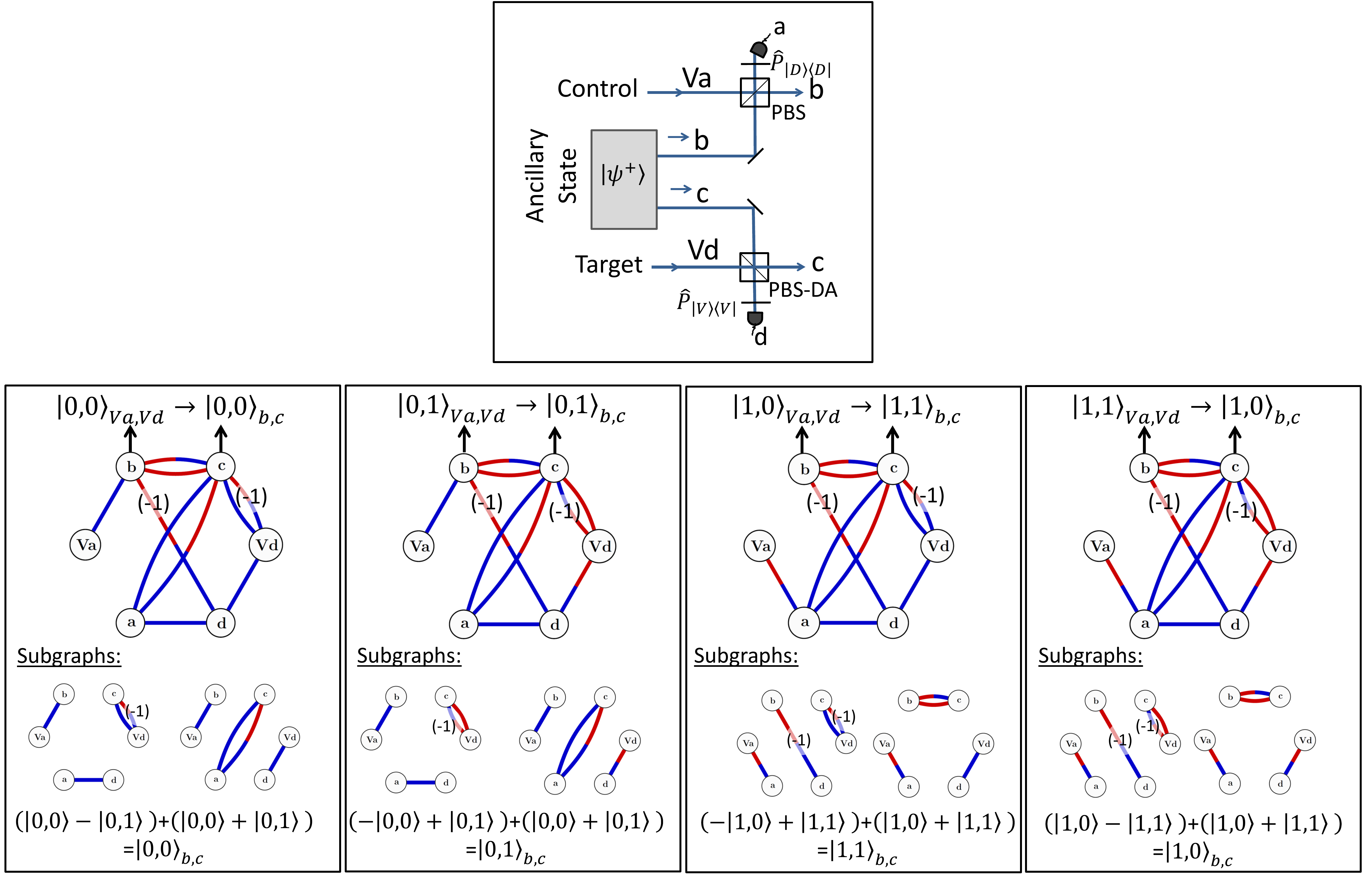}
\caption{2-dimensional \cnot gate, performed by Gasparoni et al.,\cite{gasparoni2004realization}.}
\label{figSI:CNOT2d}
\end{figure}

\begin{figure}[t]
\centering
\includegraphics[width=0.65\textwidth]{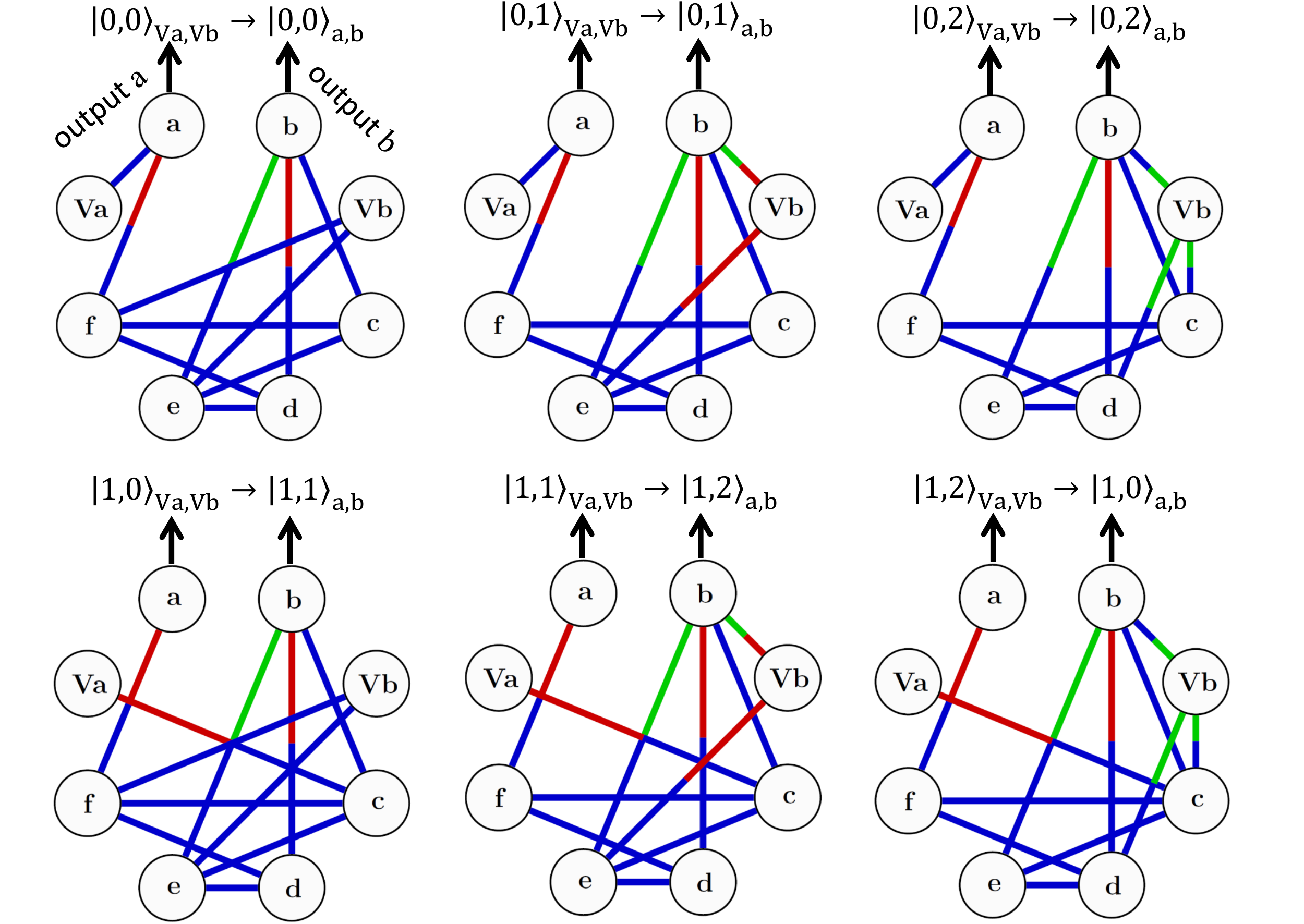}
\caption{High-dimensional \cnot gate, with a qubit control photon and a qutrit target photon.}
\label{figSI:CNOT3d}
\end{figure}

A photonic \cnot transformation was performed by Gasparoni et al.,\cite{gasparoni2004realization} in 2004, which can be seen in FigS.\ref{figSI:CNOT2d}. An ancillary state $\ket{\psi^+}=1/\sqrt{2}\left(\ket{0,1}+\ket{1,0} \right)$ in paths $b$ and $c$ is combined with the incoming control and target photons. A simultaneous detection event in detector $a$ and $d$ heralds a successful realization of a \cnot.

The corresponding graphs for the four different cases are seen below. The resulting states correspond to all subgraphs with one incoming edge in vertex $a$ and one in vertex $d$ (those are heralding detectors), and one edge from each vertex Va and Vd (those represent the incoming photons). It can be seen that Vd (which corresponds to the incoming photon from path $d$, i.e. the target photon) is responsible for the phase of the quantum states. In that way, it is responsible for the term that is destructively interfered -- this is analogous to the situation presented in the main text.

\section{CNOT beyond qubits}

A control operation in a $2\times 3$ dimensional space is shown in FigS.\ref{figSI:CNOT3d}. The subgraph $a$-$f$ remains constant, while the edges containing Va and Vb changes depending on the input control/target photons. The correct transformation is heralded by simultaneous detection of a photon in each of the detectors $c$-$f$. The structure of the subgraph $a$-$f$ is very reminiscent of the solution of heralded Bell states in Fig.4 of the main text. Here, each internal mode (represented as edge colour) from $a$ and $b$ is connected to one individual heralding detector.

Furthermore, the solution uses destructive interference for producing the correct output states, as in Fig.4 of the main text. Some of the resulting subgraphs (those have one incoming edge to vertex $c$-$f$) do not vanish. Still, they are reduced in magnitude by adapting the edge weights appropriately. Thereby, an experimentally feasible method of performing \cnot transformations beyond qubits is constructed.

\end{document}